\documentclass[fleqn,usenatbib]{mnras}

\usepackage{newtxtext,newtxmath}
\usepackage[T1]{fontenc}

\DeclareRobustCommand{\VAN}[3]{#2}
\let\VANthebibliography\thebibliography
\def\thebibliography{\DeclareRobustCommand{\VAN}[3]{##3}\VANthebibliography}
\newcommand{\ergs}{erg s$^{-1}$}
\newcommand{\ergcms}{erg s$^{-1}$ cm$^{-2}$}

\usepackage{graphicx}	% Including figure files
\usepackage{amsmath}	% Advanced maths commands
\title[Globular Cluster ULXs in NGC 1316]{Three Ultraluminous X-ray Sources Hosted by Globular Clusters in NGC 1316}

\author[K.C. Dage et al]{
Kristen C. Dage,$^{1,2}$\thanks{E-mail: kristen.dage@mail.mcgill.ca}
Arunav Kundu,$^3$
Erica Thygesen,$^{4}$
Arash Bahramian,$^{5}$
\newauthor
Daryl Haggard, $^{1,2}$
Jimmy A. Irwin,$^{6}$
Thomas J. Maccarone,$^7$
Sneha Nair,$^{1}$
\newauthor
Mark B. Peacock,$^{3}$ 
Jay Strader,$^{4}$ 
Stephen E. Zepf $^{4}$
\\
$^{1}$ Department of Physics, McGill University, 3600 University Street, Montr\'eal, QC H3A 2T8, Canada\\
$^{2}$ McGill Space Institute, McGill University, 3550 University Street, Montr\'eal, QC H3A 2A7, Canada\\
$^{3}$Eureka Scientific, Inc., 2452 Delmer Street, Suite 100 Oakland, CA 94602, USA\\
$^{4}$Department of Physics and Astronomy, Michigan State University, East Lansing, MI 48824\\
$^{5}$ International Centre for Radio Astronomy Research $--$ Curtin University, GPO Box U1987, Perth, WA 6845, Australia\\
$^{6}$Department of Physics and Astronomy, University of Alabama, Box
870324, Tuscaloosa, Alabama, 35487, USA\\
$^{7}$Department of Physics, Box 41051, Science Building, Texas Tech University, Lubbock, TX 79409-1051, USA \\
}

\date{Accepted XXX. Received YYY; in original form ZZZ}

\pubyear{2021}
\begin{document}
\label{firstpage}
\pagerange{\pageref{firstpage}--\pageref{lastpage}}
\maketitle

% Abstract of the paper
\begin{abstract}
We have identified three ultraluminous X-ray sources (ULXs) hosted by globular clusters (GCs) within NGC 1316's stellar system.  These discoveries bring the total number of known ULXs in GCs up to 20.  We find that the X-ray spectra of the three new sources do not deviate from the established pattern of spectral behaviour of the other known GC ULXs. The consistency of the X-ray spectral behaviour for these sources points to multiple paths of formation and evolution mechanisms for these rare and unique sources. Using the now larger sample of GC ULXs, we compare the optical properties of the entire known population of GC ULXs to other GCs across five galaxies and find that the properties of clusters that host ULXs are quite different from the typical clusters. Lastly, any trend of GC ULXs being preferentially hosted by metal-rich clusters is not strongly significant in this sample. 
\end{abstract}

% Select between one and six entries from the list of approved keywords.
% Don't make up new ones.
\begin{keywords}
NGC 1316: globular clusters: general -- stars: black holes -- X-rays: binaries
\end{keywords}

%%%%%%%%%%%%%%%%%%%%%%%%%%%%%%%%%%%%%%%%%%%%%%%%%%

%%%%%%%%%%%%%%%%% BODY OF PAPER %%%%%%%%%%%%%%%%%%

\section{Introduction}
Ultraluminous X-ray sources (ULXs) are off-nuclear X-ray point sources with luminosities exceeding the Eddington limit for a 10 solar mass black hole (BH), typically $\gtrsim$10$^{39}$ erg s$^{-1}$. Most ULXs tend to occur in young, star-forming regions of spiral galaxies, and a handful have identified  neutron star accretors, identified by pulsations, possibly with high magnetic fields \citep[and references therein]{2014Natur.514..202B, brightman18, 2021ApJ...909....5H}. %While many ULXs in star-forming regions of galaxies have been identified, the first ULX hosted by a globular cluster (GC) associated with older, elliptical galaxies was only identified in 2007 (Maccarone et al 2007). However, this number has been steadily increasing. 
 Statistical evidence for such sources in elliptical galaxies is poor \citep{2004ApJ...601L.143I}, but starting with the source XMMU~J122939.+075333 in the globular cluster RZ~2109 in the galaxy NGC 4472 \citep{maccarone07}, a growing number of GC ULXs in elliptical galaxies has been discovered. With the globular cluster association, the problem of distinguishing between background AGN projected against a galaxy and bona fide members of the galaxies is ameliorated, and secure identifications of ULXs in these older populations can be made.

For these GC ULXs, BH accretors offer the most likely explanation. Since these ULXs exist in crowded, older stellar populations, the only remaining donor stars are low mass and hydrogen deficient. In these dense stellar environments, binaries are formed dynamically. Given the number of differences between the old, crowded GC environment, and the young regions where NS ULXs have been identified,  this implies the accretion phenomena and compact objects of GC ULXs may be fundamentally different than the physics behind the younger NS ULXs (see also theoretical work such as \citet{2021arXiv210302026W} which suggests that the physics powering ultraluminous X-ray sources in young ($<$100  Myr) environments is vastly different than the physics in environments $>$ 2Gyr).

Identifying these BH candidates in extragalactic globular clusters is exciting because, while many studies have identified black holes and black hole candidates in Galactic globular clusters \citep{strader12, 2013ApJ...777...69C, MillerJones15,giesers18, shishkovsky18, giesers19}, the Milky Way is home to fewer than 300 globular clusters. By contrast, the total number of clusters  in even a single elliptical galaxy can be  larger than $\sim$100 the number in the Milky Way, offering tens of thousands of clusters for study. This allows for broader statistical studies of the nature of BH-hosting clusters than can be accomplished in the Milky Way.

The nature and number of these sources have important implications for the progenitors of the black hole-black hole binary mergers detected by LIGO, as globular clusters are a likely birthplace of these sources \citep{abbott16, 2019PhRvD.100d3027R}.  %In fact, new simulations by \citealt{2021arXiv210107793R} suggest that globular clusters dynamically BH-BH binaries at such a rate that can entirely explain the large number of BH-BH mergers observed by LIGO. 
Black holes are also important for understanding cluster evolution  \citep{giersz19, kremer19}, as black holes may play an important role in the cluster dynamics \citep{2020ApJ...888L..10Y}.  This marks a dramatic shift from early theoretical work, which predicted that while black holes certainly formed in globular clusters, they would be thrown out of the cluster either due to natal kicks or gravitational interactions (e.g. Spitzer 1969).

The first ULX associated with a globular cluster was identified by \citet{maccarone07}. The source displayed unusual short-term variability on the scale of hours, as well as luminous and broad [OIII] emission lines, visible above the host cluster continuum \citep{zepf08, steele11}. Analysis by \citet{peacock12neb} measured  a sizeable oxygen nebula in the system. \citet{steele14} placed limits on any excess hydrogen emission ([OIII]/H$\beta$ =106) and identified a carbon-oxygen white dwarf as the very likely donor star of the system. The long-term X-ray and optical properties of the source were studied by \citet{dage18, dage19b}, which showed that the X-ray luminosity varied by several orders of magnitude over 16 years, and the optical luminosity of the line declined over a 7 year period. Subsequent observations indicate that the optical luminosity has been rising for the last two years, placing constraints on the size of the oxygen nebula (between $10^{-3}$ pc at minimum, but possibly up to 2pc). 

In 2010, two GC ULXs were identified in NGC 1399, the central galaxy in the Fornax Cluster. One of these showed short variability on the timescales of hours, and was bright in X-rays until 2003, but has not been detected since  \citep{shih10}. The other GC ULX did not show variability within an observation, but showed narrow [OIII] and [NII] emission lines above the host cluster continuum \citep{irwin10}.

\citet{maccarone11} discovered a second GC ULX hosted by one of NGC 4472's clusters, and \citet{roberts12} identified  a GC ULX in an NGC 4649 cluster. Optical spectroscopy of the NGC 4649 cluster revealed no emission lines beyond the host cluster continuum, implying that not every GC ULX is host to optical emission.  \citet{Irwin16} discovered X-ray sources hosted by two globular clusters that flared above the Eddington limit on the timescale of just minutes.  Similar sources were also identified by \citet{2005ApJ...624L..17S}.

\citet{dage19a} conducted the first ever large-scale, long-term X-ray analysis of GC ULXs in NGC 4472, NGC 4649 and NGC 1399, comparing the X-ray spectral properties of these and other sources. One of the major results of this work was that the two clusters that were known to exhibit optical emission lines showed markedly different X-ray behaviour than the rest of the sample, with lower best-fit inner disk temperatures and spectral shapes uncorrelated with major changes in X-ray luminosity. 

Seven more GC ULXs were identified by \citet{dage20}, associated with the large GC population of M87. The X-ray spectra of these sources behaved like the larger GC ULX sample (and unlike the optical emission line GC ULXs). A third source with significant short-term variability was identified. Lastly, it was shown that the ULXs were preferentially hosted by brighter (and hence, presumably, more massive) clusters, but did not show evidence for a correlation between the cluster colour and presence of a ULX.

In this work, we look to NGC 1316 to identify new GC ULX sources. NGC 1316 is a giant early-type galaxy that underwent a merger as recently as 1--3Gyr ago  (\citealt{goudfrooijmerger}, see also work by \citet{2021arXiv210110347K} for a discussion of the galaxy's history). Studies of the galaxy's globular cluster system show that it is made up of two distinct populations of younger and older clusters \citep{goudfrooi01}. Studies of the HI content of the galaxy give further agreement that NGC 1316 is the product of a merger of a lenticular and spiral galaxy \citep{2021arXiv210110347K} .  
 
 We use archival \textit{Chandra} and HST data to identify three new GC ULXs in NGC 1316, and compare the X-ray spectral properties of the sources to the previously-identified GC ULXs, as well as the optical properties of the host clusters. Section \ref{sec:obs} describes the data and analysis techniques used for both optical and X-ray data reduction. Section \ref{sec:res} discusses the results of this work and compares the X-ray and optical properties of the 20 known GC ULXs. Finally, the implications of this analysis are discussed in Section \ref{sec:conc}. 
 
\section{Observations and Analysis}
\label{sec:obs}

\begin{figure*}
%\begin{tabular}{ll}
\includegraphics[width=6.in]{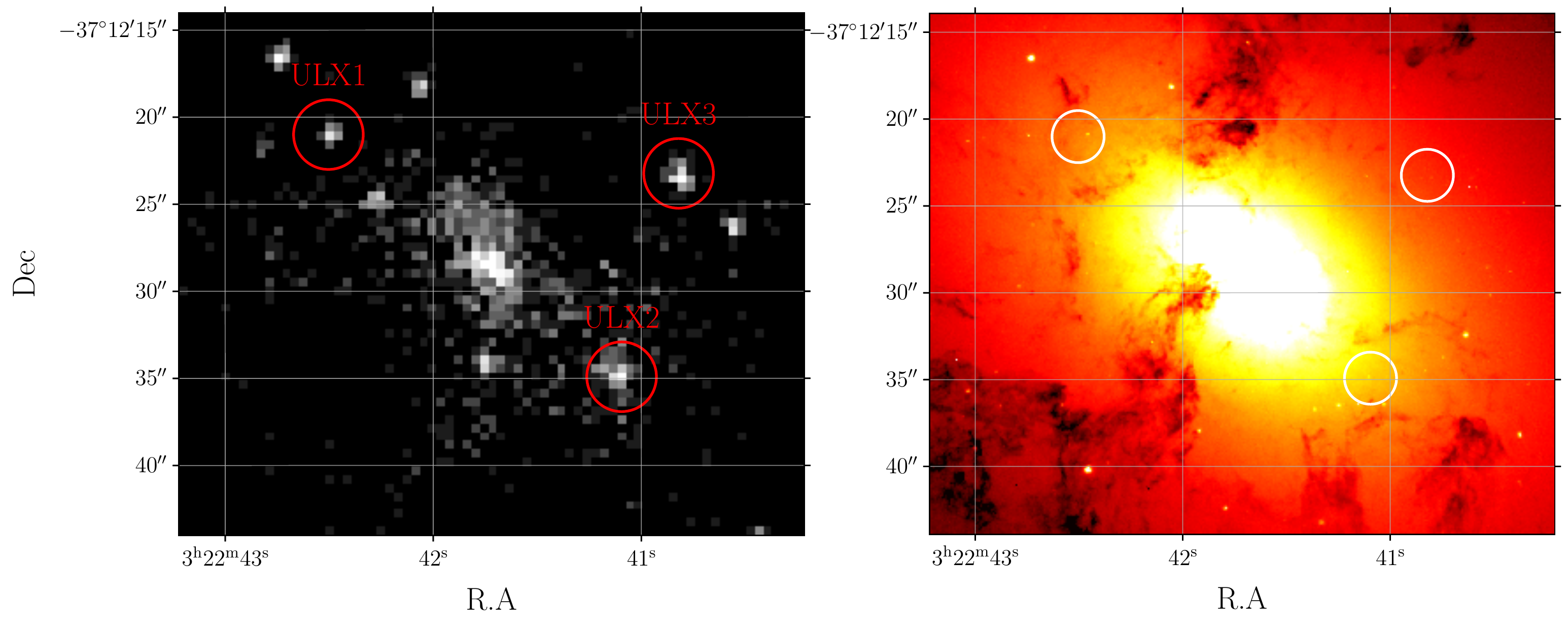}
%&
%\includegraphics[width=3.in, angle =90]{gimg.png}
%\end{tabular}
\caption{Left: X-ray image of NGC 1316 (ObsID 20341) with positions of the three GC ULXs overlaid. Right: HST F475W (Sloan $g$ band) image of NGC 1316 with the GC ULXs overlaid.   } 
\label{Fig:322_321}
\end{figure*}
We analysed archival X-ray and optical data of NGC 1316 to identify three ULXs hosted by globular clusters in this galaxy (see Figure \ref{Fig:322_321}). The data are displayed in Table \ref{chandradata}, and include six \textit{Chandra} X-ray observations, with one taken in 2001 and five spread across a single month in 2019.  We use HST observations in the F475W and F850LP filters (Sloan $g$ and $z$ bands in the AB magnitude system) to identify potential globular cluster counterparts, as described in Section \ref{sec:hst}.

\subsection{Identification of ULX sources}
To identify the X-ray point source populations, we use the function \textsc{wavdetect} from \textsc{ciao} on each X-ray image, which were cleaned and filtered of background flares, with the exposure map centered at 2.3 keV, and wavelet scales of 1.0, 2.0, 4.0, 8.0 and 16.0 pixels. We use an enclosed count fraction (e.c.f.) of 0.3, and significance threshold
of $10^{-6}$.  This corresponds to about one false detection per chip, although the false detection will not have a flux corresponding to a ULX. We then estimate the fluxes of the identified sources by using the function \textsc{srcflux}, assuming a fixed power-law of $\Gamma$=1.7, the typical power-law index for X-ray binaries, and fixed the hydrogen column density, $N_H$, to the line of sight density, 2.13 $\times 10^{20}$ cm$^{-2}$ \citep{1990ARA&A..28..215D}. We use the distance of 20.0 Mpc \footnote{The distance to the Fornax cluster used in \citet{dage19a}.} to estimate the X-ray luminosity for all of the sources. This yields three unique X-ray point sources with luminosities near or exceeding the Eddington limit. While this analysis is only concerned with sources that have X-ray luminosities exceeding the Eddington limit for a 10 solar mass BH, it is worth noting that there are a number of less bright X-ray sources associated with globular clusters in the sample from \citet{jordan09}.

\begin{table*}
\centering
\caption{Left column: \textit{Chandra} observations of NGC 1316, all observations are ACIS-S. Right column: \textit{HST} observations for $g$ and $z$ (AB mag) observations.}
\label{chandradata}
\begin{tabular}{llllllll}
\hline \hline
\multicolumn{1}{|l|}{ObsID (\textit{Chandra})} & \multicolumn{1}{l|}{Date} & \multicolumn{1}{l|}{Exposure (ks)} & ObsID (\textit{HST})& Date& Filter  & Exposure (s) \\ \hline
 2022 & 2001-04-17 & 30 & j90x01020 & 2005-02-16&F475W ($g$) & 760\\
 20340 & 2019-04-10& 45 &j90x01010 & 2005-02-16&F850LP ($z$) &1130 \\
 20341 & 2019-04-22 & 52\\
 22179 & 2019-04-17& 39\\
 22180 & 2019-04-20 & 14\\
 22187 & 2019-04-25 & 53\\
\hline
\end{tabular}
\end{table*}

\subsection{HST analysis and globular cluster counterpart identification}
\label{sec:hst}
We use catalogs of the globular cluster population of NGC 1316 from  \citet{goudfrooi01} and \citet{jordan16} to identify three new ultraluminous X-ray sources hosted by globular clusters. Two ULX sources, GCULX1 and GCULX2, match  to globular cluster candidates in the \citet{jordan16} survey. GCULX3's optical counterpart, identified in \citet{allak}, also had characteristics typical of extragalactic globular clusters. We use  DAOPHOT \citep{1987PASP...99..191S} to perform photometry on all three sources, as well as other known GCs in the field, using images that corresponds  to the $g$, $z$ bands, and found that GCULX3's colour and magnitude are typical for a globular cluster. The ACS and WFC3 zero-points adopted from the HST ACS and WFC3 data handbook \footnote{\url{https://www.stsci.edu/hst/documentation}}.

We measure the half-light radius with BAOlab \citep{larsen99} using the KING30 model convolved with a PSF made in Tiny Tim \citep{2011SPIE.8127E..0JK}, and found the size of GCULX3 was on par with the globular clusters identified as having a 100\% probability of being a globular cluster in \citet{jordan16}. 

\begin{figure}

\includegraphics[width=3.5in]{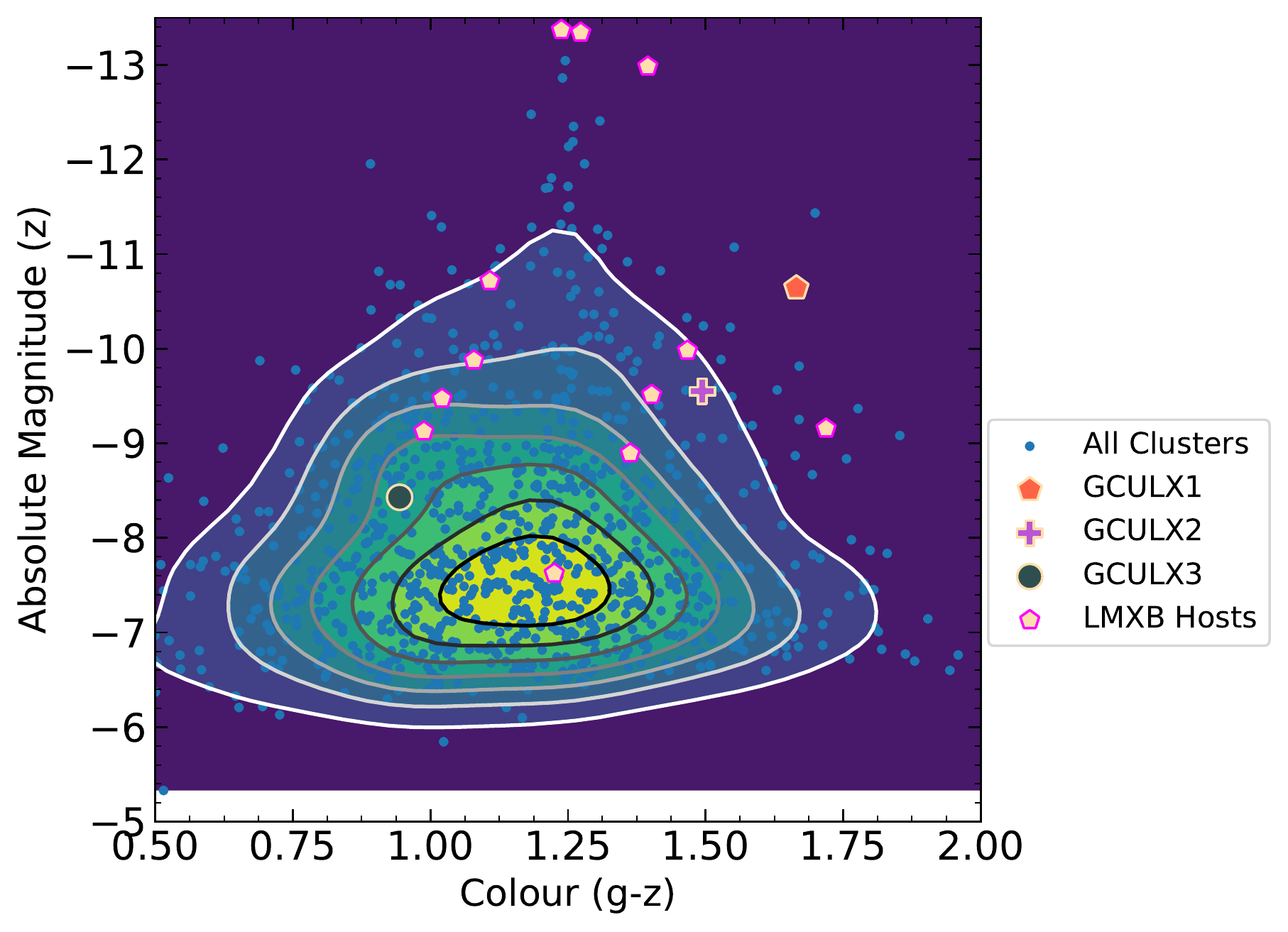}

\caption{Colour and magnitude of the three GC ULXs in NGC 1316 compared to the overall population of globular clusters.} 
\label{fig:zgz}
\end{figure}

The coordinates of the three GC ULX sources are listed in Table \ref{sources}, along with their measured colour and magnitude in the AB system. The optical properties of the host clusters compared to the rest of the cluster population in the NGC 1316 are displayed in Figure \ref{fig:zgz}.

\begin{table*}
\caption{X-ray coordinates of the three ULXs and lower X-ray luminosity sources coincident with candidate globular clusters. These sources have been identified in \citet{2003ApJ...586..826K}, \citet{2004ApJS..154..519S}, and \citet{allak}.  Sources marked with $^o$ were identified by \citet{goudfrooi01}, sources with $\dagger$ were identified by \citet{jordan16}.   X-ray luminosity estimates are based off of \textsc{srcflux} calculations from ObsID 22187.  Source counts and errors are based off of \textsc{wavdetect} measurements of ObsID 22187.} CXOU J032238-371228 was only measured in the Vega magnitude system by \citet{goudfrooi01}, and hence there are no AB magnitudes to report.
\label{sources}
\begin{tabular}{cclccccccc}
 \hline \hline
 Name      & RA           & Dec          & $L_X$ Estimate & Counts   & z-band  &$g-z$  \\ 
           &              &              & $\times 10^{39}$\ergs &0.5-7.0 keV& (AB mag) & \\
 \hline
CXOU J032242.5-371222 (GCULX1) $\dagger$ & 03:22:42.48 &-37:12:21.15 &1.39&112 $\pm$ 11& 20.88 &1.67 \\
CXOU J032241.07-371235.3 (GCULX2) $\dagger$  & 03:22:41.11 & -37:12:34.73 &     1.05&61 $\pm$ 9 &21.98 & 1.49&\\
CXOU J032240.8-371224 (GCULX3)  & 03:22:40.82 &-37:12:23.28 &2.15& 159 $\pm$ 16& 23.10 &0.94\\   

CXOU J032244-371310 $^o$ & 03:22:44.71& -37:13:10.07& 0.15& 9 $\pm$3& 20.81 & 1.11\\
CXOU J032241-371304 $^o$ $\dagger$ & 03:22:41.89& -37:13:04.49& 0.33 &17 $\pm$4 & 22.64&  1.36\\
CXOU J032242-371258 $^o$ & 03:22:42.43& -37:12:58.83& 0.85 &61 $\pm$8& 22.37 & 1.72\\
CXOU J032240-371244 $^o$ & 03:22:40.42& -37:12:44.51& 0.13&16 $\pm$5& 22.02 & 1.40\\
CXOU J032237-371251 $^o$ & 03:22:37.69& -37:12:51.11& 0.21 &16 $\pm$4& 18.19 & 1.27\\
CXOU J032238-371211 $^o$ & 03:22:38.19& -37:12:11.58& 0.12 &9 $\pm$3& 22.06&  1.02\\
CXOU J032239-371148 $^o$  $\dagger$ & 03:22:39.12& -37:11:48.00& 0.70 &64 $\pm$8& 21.55 & 1.47\\
CXOU J032242-371218 $^o$ & 03:22:42.06& -37:12:18.25& 0.41 &35 $\pm$7& 18.55 & 1.40\\
CXOU J032242-371216 $^o$ $\dagger$ & 03:22:42.73& -37:12:16.62& 0.41 &36 $\pm$6& 18.16 & 1.24\\
CXOU J032242-371206 $^o$  $\dagger$ & 03:22:42.60& -37:12:06.25& 0.20 &10 $\pm$3& 22.40 & 0.99\\
CXOU J032242-371123  $^o$ & 03:22:42.10& -37:11:23.92& 0.42 &47 $\pm$7& 23.91 & 1.23\\
CXOU J032241-371117 $^o$ $\dagger$ & 03:22:41.29& -37:11:17.13& 0.30 &19 $\pm$4& 21.65 & 1.08\\
CXOU J032238-3712 $\dagger$ & 03:22:38.78& -37:12:28.66& 0.13&17 $\pm$4& N/A & N/A\\
\end{tabular}
\end{table*}

\subsection{X-ray Analysis}

\begin{table*}
\caption{Total background subtracted counts in the 0.5-8.0keV band and best fit spectral parameters for combined spectra of the three GC-ULXs. }
\label{deepfits}

\begin{tabular}{llcccccccc}  \hline \hline
          & & \multicolumn{3}{c}{power-law}                                &&  \multicolumn{3}{c}{diskbb}   \\    
\cline{3-5}\cline{7-9}\\
Source     &Total Counts (0.5-8.0 keV) & $\Gamma_\text{PL}$ & $\chi_\nu^2$/d.o.f. & F-Test prob. (\%) && kT$_{\text{in}}$ (keV)   & $\chi_\nu^2$/d.o.f.  & F-Test prob. (\%) \\ 
\hline 
GCULX1   &272   & 1.4 ($\pm 0.3$)&    0.42/11         & --                        & &1.5 ($^{+0.7}_{-0.4}$)&   0.62/11    &      --      &  \\
GCULX2 &613& 1.8 ($\pm$0.1) &     1.29/26    &5&&    1.2  ($\pm$0.2)   &  1.08/26         &  53
  \\
GCULX3 &484 &1.6($\pm 0.1$)  & 1.01/21           & ---                       && 1.3 ($\pm$0.2)  &    0.87/21    & --  \\

\hline    
\end{tabular}
\end{table*}

\begin{table*}
\centering
    \caption{\textit{Chandra} Fit Parameters and Fluxes (0.5-8 keV) for spectral best fit single-component models,  \texttt{tbabs*pegpwrlw} and\texttt{tbabs*diskbb} for NGC 1316-GCULX1. Hydrogen column density ($N_H$) frozen to 2.13 $\times 10^{20}$cm$^{-2}$. All fluxes shown are unabsorbed in the 0.5-8 keV band. In ObsID 22180, GCULX1 has a background subtracted count rate of 8.84 $\times 10^{-4}$ counts/sec in the 0.5-8 keV range. The upper limit was calculated using a fixed power-law index of $\Gamma$ = 1.4. }
\label{ulxfits-ulx1}
\begin{tabular}{ccccccccc}
\hline
\hline
\multicolumn{9}{c}{NGC 1316 GCULX1}\\ 
\hline
                    & \multicolumn{3}{c}{\textsc{tbabs*pegpwrlw}} && \multicolumn{4}{c}{\textsc{tbabs*diskbb}} \\ 
\cline{2-4}\cline{6-9}\\
ObsID (Date)       & $\Gamma$           &   $\chi^2_{\nu}$/d.o.f. & PL Flux && $T_{in}$ &  Disk Norm &  $\chi^2_{\nu}$/d.o.f. &  Disk Flux  \\
                   &                    &               &($10^{-14}$ \ergcms)&& (keV)&($10^{-4}$)&&($10^{-14}$ \ergcms)\\ 
\hline
 2022 (2001-04-17) & 1.3 ($\pm$0.4)  &   (44.36/51)     & 2.9 ($^{+1.2}_{-0.9}$) &&  1.7 $(^{+2.2}_{-0.6})$ &$\leq$1.7& (44.44/51)&2.4 ($^{+1.3}_{-0.7}$)\\ 
20340 (2019-04-10)& 1.2 ($\pm$ 0.5) &   (70.61/63)     & 1.7 ($\pm$0.6)  && 2.0 ($^{+4.5}_{-0.8}$)  &  $\leq$0.5& (69.02/63) &  1.5 ($\pm$ 0.6) \\
20341 (2019-04-22)& 1.6 ($\pm$0.6)  &   (30.78/34)     & 1.3 ($\pm$0.5) &&1.5 ($^{+1.8}_{-0.6}$) &$\leq$  1.1&(31.62/34) &1.1 ($\pm$ 0.5)\\
22179 (2019-04-17)& 1.9 ($\pm$0.7)  &      (32.05/30)  & 1.2($\pm$0.5)  &&  1.4($^{+2.3}_{-0.6}$)  &$\leq$  1.3    &(35.09/30)&  1.0 ($\pm$ 0.5)\\
22180 (2019-04-20)& $1.4$  &      -  & $\leq$ 1.3  && -&- &-&- \\ 
22187 (2019-04-25)& 1.4 $(\pm 0.4)$ &   (47.22/60)      & 1.9 ($\pm$ 0.5) && 1.6 ($^{+1.3}_{-0.5}$ &$\leq$  1.2& (47.74/60)& 1.7 ($\pm$ 0.6) \\ \hline
\end{tabular}
\label{ulx1fits}
\end{table*}

\begin{table*}
\centering
    \caption{\textit{Chandra} Fit Parameters and Fluxes (0.5-8 keV) for spectral best fit single-component models,  \texttt{tbabs*pegpwrlw} and\texttt{tbabs*diskbb} for NGC 1316-GCULX2. Hydrogen column density ($N_H$) frozen to 2.13 $\times 10^{20}$cm$^{-2}$. All fluxes shown are unabsorbed in the 0.5-8 keV band. }
\label{ulxfits-ulx2}
\begin{tabular}{ccccccccc}
\hline
\hline
\multicolumn{9}{c}{NGC 1316 GCULX2}\\ 
\hline
                    & \multicolumn{3}{c}{\textsc{tbabs*pegpwrlw}} && \multicolumn{4}{c}{\textsc{tbabs*diskbb}} \\ 
\cline{2-4}\cline{6-9}\\
ObsID (Date)       & $\Gamma$           &   $\chi^2_{\nu}$/d.o.f. & PL Flux && $T_{in}$ &  Disk Norm &  $\chi^2_{\nu}$/d.o.f. &  Disk Flux  \\
                   &                    &               &($10^{-14}$ \ergcms)&& (keV)&($10^{-3}$)&&($10^{-14}$ \ergcms)\\ 
\hline
 2022 (2001-04-17) & 1.6 ($\pm$0.7)  &  0.362/3      & 3.3 ($^{+3.0}_{-1.3}$) &&  0.7 $(^{+1.6}_{-0.25})$ &$\leq$3.7& 0.23/3&1.7 ($^{+2.4}_{-0.5}$)\\ 
20340 (2019-04-10)& 1.5 ($\pm$ 0.5) &  1.139/61      & 2.2 ($\pm$0.6)  && 1.4 ($^{+0.9}_{-0.4}$)  &$\leq$0.2 &   0.9385/61&  1.8 ($\pm$ 0.5) \\
20341 (2019-04-22)& 1.4 ($\pm$0.6)  &  0.848/3       & 3.1 ($^{+1.2}_{-0.9}$) &&1.2 ($^{+1.8}_{-0.4}$) &  $\leq$ 0.5 &0.561/3 &2.2($^{+1.3}_{-0.6}$)\\
22179 (2019-04-17)& 2.2($\pm$0.6)  &    (65.04/52)    & 2.0($\pm$0.5)  &&  0.9($^{+0.5}_{-0.3}$)  &$\leq$ 1.6&(  69.92 /52)&  1.5 ($\pm$ 0.5)\\
22180 (2019-04-20)& 1.9 ($\pm$ 0.8)  &    (38.26/29)    & 2.7 ($^{+1.2}_{-0.9}$)  && 1.1($^{+1.6}_{-0.4}$)  &$\leq$0.7 &(38.15/29)&2.1 ($^{+1.2}_{-0.8}$)  \\ 
22187 (2019-04-25)& 1.7 $(\pm 0.4)$ &     0.93/5     & 3.7($\pm$ 0.6) && 1.2 ($^{+0.6}_{-0.3}$ &$\leq$ 0.7& 1.29/5& 2.9 ($\pm$ 0.6) \\ \hline
\end{tabular}
\label{ulx2fits}
\end{table*}

\begin{table*}
\centering
    \caption{\textit{Chandra} Fit Parameters and Fluxes (0.5-8 keV) for spectral best fit single-component models,  \texttt{tbabs*pegpwrlw} and\texttt{tbabs*diskbb} for NGC 1316-GCULX3. Hydrogen column density ($N_H$) frozen to 2.13 $\times 10^{20}$cm$^{-2}$. All fluxes shown are unabsorbed in the 0.5-8 keV band. In ObsID 22180, the upper limit for GCULX3 was estimated using a background subtracted count rate of 9.58 $\times 10^{-4}$ counts/sec in the 0.5-8 keV range.}
\label{ulxfits-ulx3}
\begin{tabular}{ccccccccc}
\hline
\hline
\multicolumn{9}{c}{NGC 1316 GCULX3}\\ 
\hline
                    & \multicolumn{3}{c}{\textsc{tbabs*pegpwrlw}} && \multicolumn{4}{c}{\textsc{tbabs*diskbb}} \\ 
\cline{2-4}\cline{6-9}\\
ObsID (Date)       & $\Gamma$           &   $\chi^2_{\nu}$/d.o.f. & PL Flux && $T_{in}$ &  Disk Norm &  $\chi^2_{\nu}$/d.o.f. &  Disk Flux  \\
                   &                    &               &($10^{-14}$ \ergcms)&& (keV)&($10^{-3}$)&&($10^{-14}$ \ergcms)\\ 
\hline
 2022 (2001-04-17) & 1.2 ($\pm$0.5)  &  0.08/2   & 1.7 ($\pm$ 0.6) &&  0.8 $(^{+0.5}_{-0.3})$ &$\leq$ 1.6&0.26/2 &1.2($\pm$ 0.5)\\ 
20340 (2019-04-10)& 1.7 ($\pm$ 0.5) &   ( 46.78/48)     & 1.7 ($\pm$0.7)  && 1.2 ($^{+0.9}_{-0.4}$)  &$\leq$0.3& (47.78/48)&  1.4 ($\pm$ 0.5) \\
20341 (2019-04-22)& 1.6 ($\pm$0.5)  &  1.16/3      & 3.4 ($^{+1.2}_{-0.5}$) &&1.1 ($^{+1.0}_{-0.3}$) &$\leq$0.9& 0.98/3 &2.3 ($^{+1.1}_{-0.6}$)\\
22179 (2019-04-17)& 1.6 ($\pm$0.6)  &    (34.44/51)    & 2.3($\pm$0.7)  &&  1.4($^{+1.9}_{-0.3}$)  &$\leq$0.3&(34.48/51)&  1.9 ($^{+0.8}_{-0.5}$)\\
22180 (2019-04-20)& $1.6$  &   --     & $\leq$ 1.7  && -&- &-&- \\ 
22187 (2019-04-25)& 1.2 ($\pm 0.5$))&   0.34/3       & 4.5($^{+1.4}_{-1.1}$) && 2.1 ($^{+2.1}_{-0.9}$ &$\leq$0.1&0.422/3 & 3.7($^{+2.1}_{-1.1}$) \\ \hline
\end{tabular}
\label{ulx3fits}
\end{table*}

 These sources were analysed in the same manner as \citet{dage19a, dage20}, in order to compare the X-ray properties of all of the 20 identified GC ULXs to each other. The sources were extracted in each observation using \textsc{ciao}, and the background regions were created by placing regions of a similar size to the source region near the source, but avoiding any nearby point sources. We use \textsc{xspec} \citep{Arnaud96} to fit the spectra in the 0.5-8.0 keV band with single component models: \texttt{tbabs*diskbb} and \texttt{tbabs*pegpwrlw}. Any observation with greater than 100 counts is binned by 20, and any observation with fewer than 100 counts is binned by 1 and fit using the Cash statistic \citep{1979ApJ...228..939C}. In ObsID 22180, both GCULX2 and GCULX3 have too few counts to extract a spectrum, and we estimate an upper limit using \textsc{pimms}. The X-ray fits for each source can be found in Tables \ref{ulxfits-ulx1}-\ref{ulxfits-ulx3}. Finally, we use \textsc{combine\_spectra} to combine all of the extracted spectra for each source (see Table \ref{deepfits} for the best-fit values). The best fit power-law index of the combined spectra was used to estimate the upper limits for GCULX2 and GCULX3 in ObsID 22180. We also fit the two component model \texttt{tbabs*(diskbb+pegpwrlw)} to the deep spectra, however, only GCULX2 could be fit by the two component model. We use F-test to determine whether the two component model was statistically a better fit than either single component model, but given that the F-test null hypothesis probability was high for either case, this implies that the complexity of the two-component model is not justified. Since this is likely due to the source having many more counts than the other two sources, we do not report the values.
 
Lastly, we extract background subtracted light-curves using \textsc{dmextract} and find no evidence of strong variability (see also analysis by \citet{allak} for lack of variability in GCULX3). This is typical for most GCULXs, currently only three of them show strong variability (see \citet{maccarone07, shih10, dage20} for more details). 
\section{Results}
\label{sec:res}
We identified three new GC ULX candidates in NGC 1316, using archival \textit{Chandra} and HST observations. The HST observations revealed three globular cluster candidates with $z$ and $g-z$, which all had similar sizes. We extracted and fit the X-ray spectra of each source across the six \textit{Chandra} observations. We fit two different single component models to the individual spectra, and also fit models to the combined spectra. Of these, only GCULX2 was able to be fit by a two component model over a single component model\footnote{and it is worth noting that GCULX2 has significantly more counts than either of the other two sources}, but in either case the F-test probability was high, meaning that it is not statistically reasonable to add the extra component. Very few GCULXs can be well-fit by a two-component model, only two known sources (RZ2109 \citep{dage18}, and M87-GCULX1 \citep{dage20} have any evidence for the extra component statistically being a better fit. 

\begin{figure}

\includegraphics[width=3.5in]{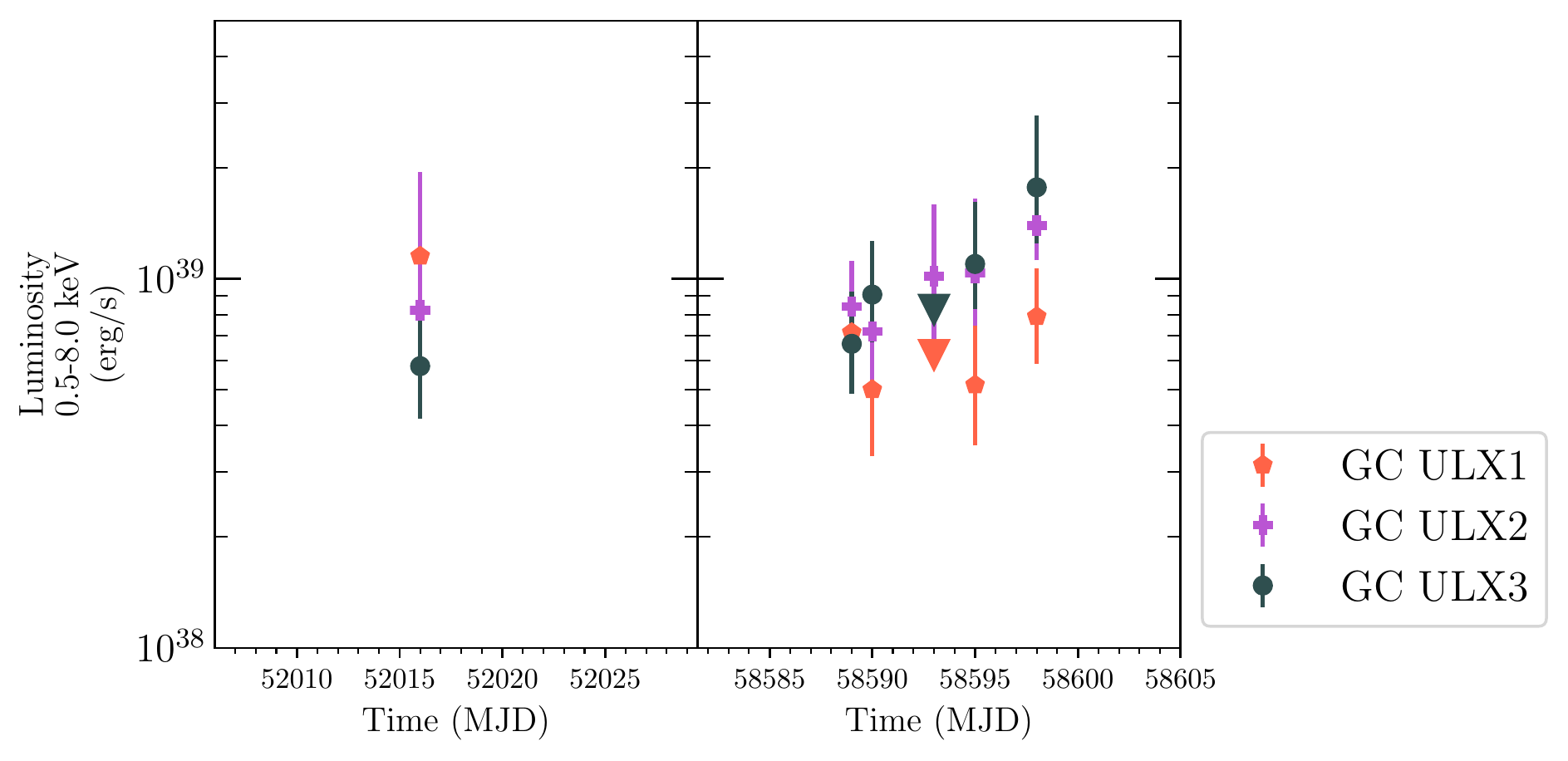}

\caption{Long-term X-ray luminosity for the three GC ULX sources. Left panel shows the observation from April, 2001 (ObsID 2022), and the right panel shows the five observations taken in April, 2019, including upper limits from ObsID 22180 (triangles).} 
\label{fig:var}
\end{figure}

Both on the one month and the 18 year time baselines probed, there is no evidence for variations for more than a factor of $\sim$ 2, which is consistent with most of the known GC ULXs \citep{dage19a}.  Only the source in RZ~2109 in NGC 4472 \citep{maccarone07}, the faded GC ULX identified by \citet{shih10} in NGC 1399, and the source SC302 in M87 \citep{dage20} show strong variability on the timescale of hours. Given that most of the observations are clustered within a single month in 2019, with just one observation occurring in 2001 (see Figure \ref{fig:var}), it is difficult to characterise any long-term variability of these sources. 

We also identified thirteen X-ray sources coincident with GCs (Table \ref{sources}), with estimated X-ray luminosities on the order of $10^{38}$ erg/s. Given the paucity of the data, we do not perform spectral fits, however, a comprehensive study of similar sources has previously been studied in \citet{2003ApJ...587..356I}. 

\subsection{Host Cluster Colour and Magnitude}
\begin{figure*}
%\begin{tabular}{ll}
\includegraphics[scale=0.6]{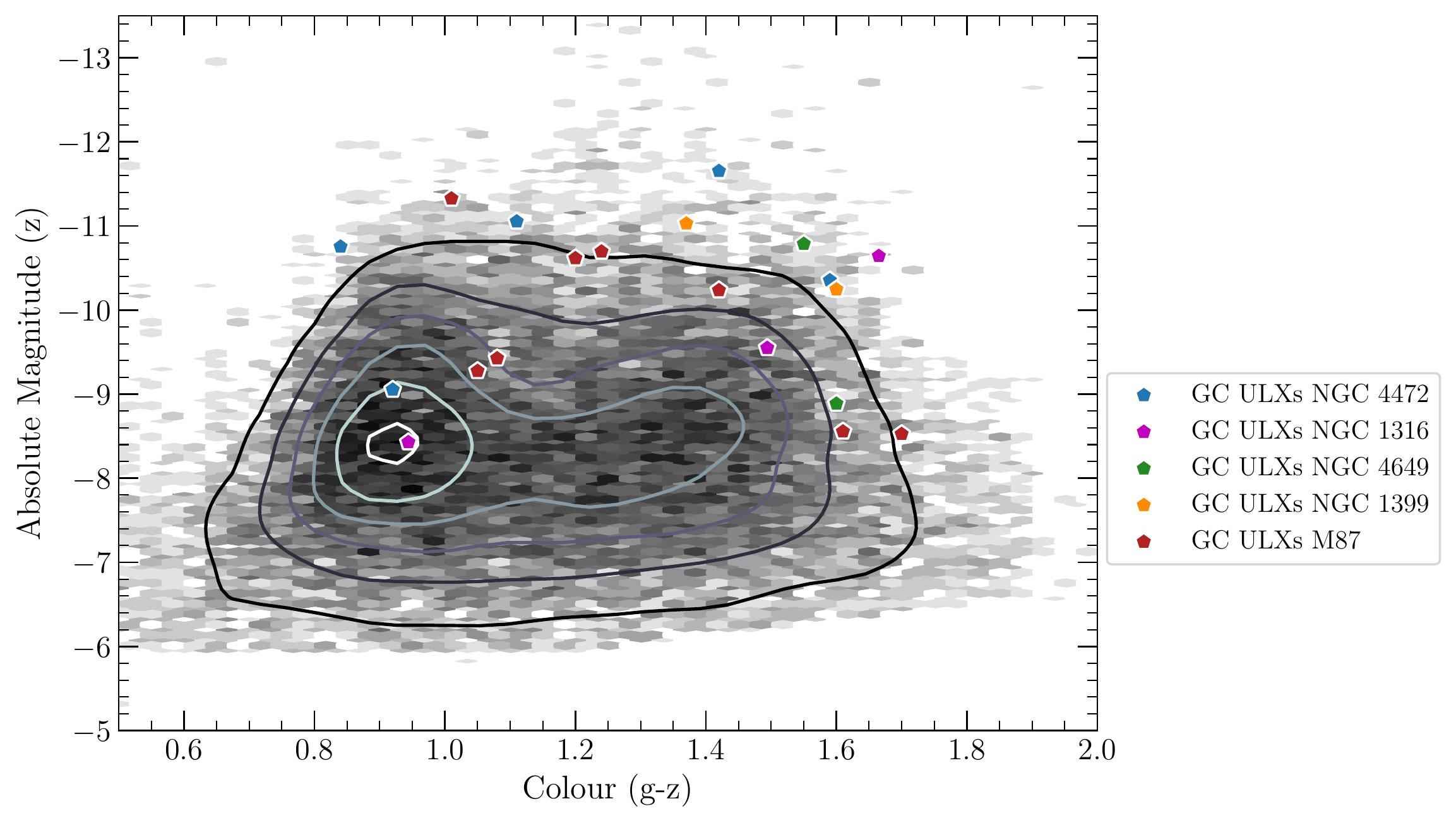}
%&
%\includegraphics[scale=0.4]{Dists.pdf}
%\end{tabular}
\caption{Colour versus Magnitude (AB, dereddened) for the globular cluster ULXs, superimposed on the properties of non-ULX hosting clusters in the five galaxies (NGC 1316, NGC 1399, NGC 4472, NGC 4649 and M87, using data from \citet{jordan09,2012ApJ...760...87S, peacock14,jordan16}). } 
\label{fig:allulxs}
\end{figure*}

In both \citet{dage19a} and \citet{dage20}, we noted that while the ULXs are indeed preferentially hosted by brighter clusters, there is not yet evidence that ULXs hold an affinity for metal rich clusters (unlike the rest of the less-luminous X-ray binary population which shows red clusters outnumbering bluer clusters by three to one as LMXB hosts \citealt{kundu02, sarazin03}). Initial studies on a low population of ULX hosting clusters by \citet{maccarone11} suggested that it was likely that ULXs are also preferentially hosted by metal rich clusters, however the sample size in this initial study was very small. With the new, larger sample of ULXs hosted by globular clusters, this question can now be revisited.%
Nineteen of the GC ULXs have colour and magnitudes measured in the AB magnitude system (see \citet{dage19a,dage20} for more). As seen in Figure \ref{fig:allulxs}, the majority of ULXs are not hosted by typical globular clusters for these galaxies, which certainly highlights the uniqueness of these already-rare sources. We used the Anderson Darling test \citep{Anderson1952,Anderson1954} from \textsc{kSamples}\footnote{\url{https://cran.r-project.org/web/packages/kSamples/index.html}} to independently compare the optical colour and magnitude distributions of the ULX hosts and other globular clusters. For the magnitude, the statistic is 22.1 and the significance level is $5.9\times10^{-10}$. For colour, the statistic value is 3.22 and the significance level is 0.016. The low significance levels imply that the magnitude of the ULX hosts come from very different distributions than the rest of the clusters in these galaxies. However, the tests cannot yet reject the null hypothesis concerning the colour distribution.% 

NGC 1316 is known to host an intermediate age population, and therefore the optical colours might not be linearly correlated metallicity for some of NGC 1316's GCs. Infrared colours are known to be more sensitive to metallicity \citep{2007ApJ...660L.109K}. To check the metallicities of the NGC 1316 ULX hosting cluster, we also performed photometry in the H-band (F160W, HST observation ib3n03050). GCULX1 was $H$=20.0$\pm$0.2, GCULX2 was $H$= 20.1$\pm$0.4  and GCULX3 was $H$= 21.3$\pm$0.4. The very red infrared colours (z-H) imply that all three ULX hosting clusters of NGC 1316 may be very metal rich. The apparent bluer optical colour of GCULX3 might be because these GCs may be from an intermediate age population \citep{goudfrooi01}. %

\begin{figure*}
\begin{tabular}{ll}
\includegraphics[scale=0.4]{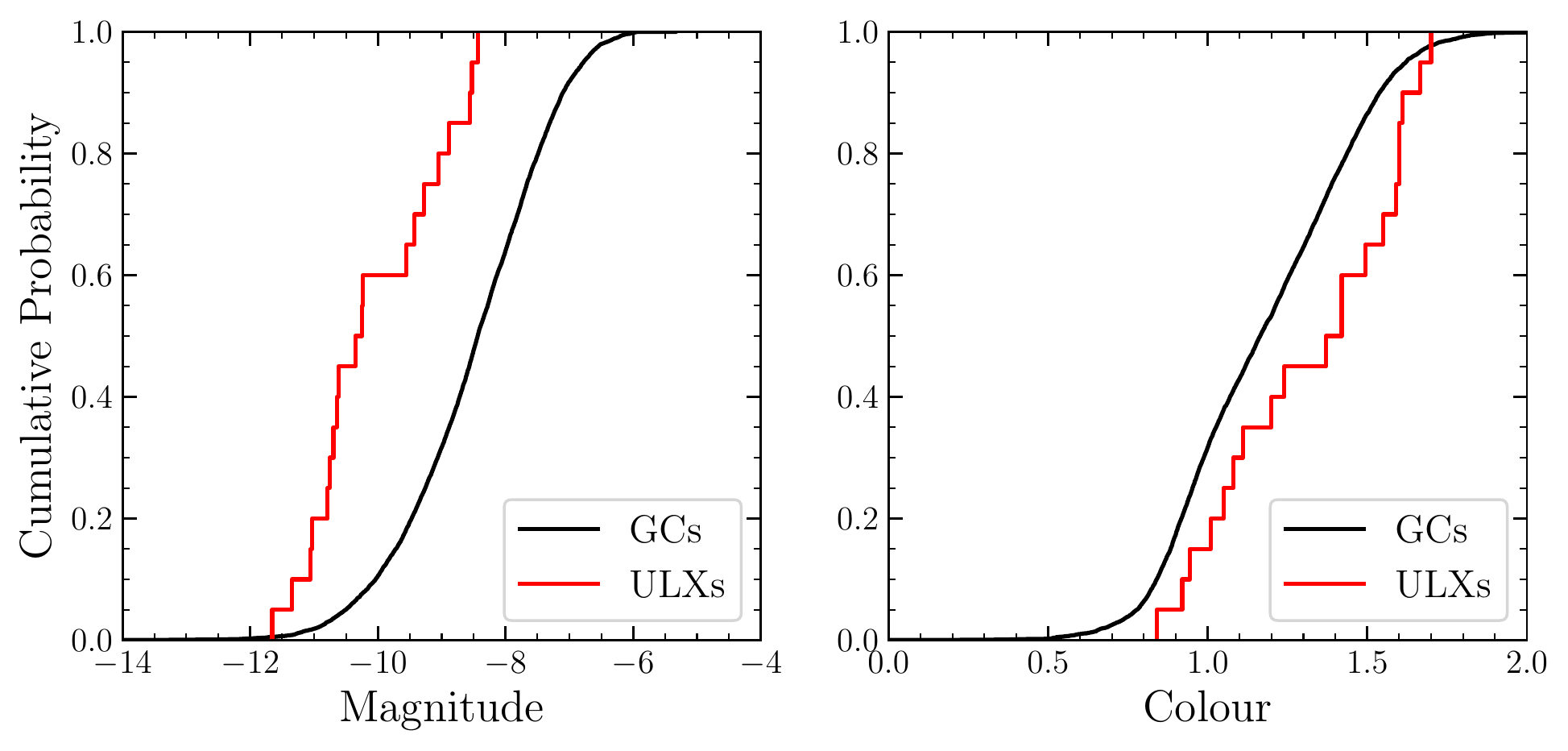}
&
\includegraphics[scale=0.4]{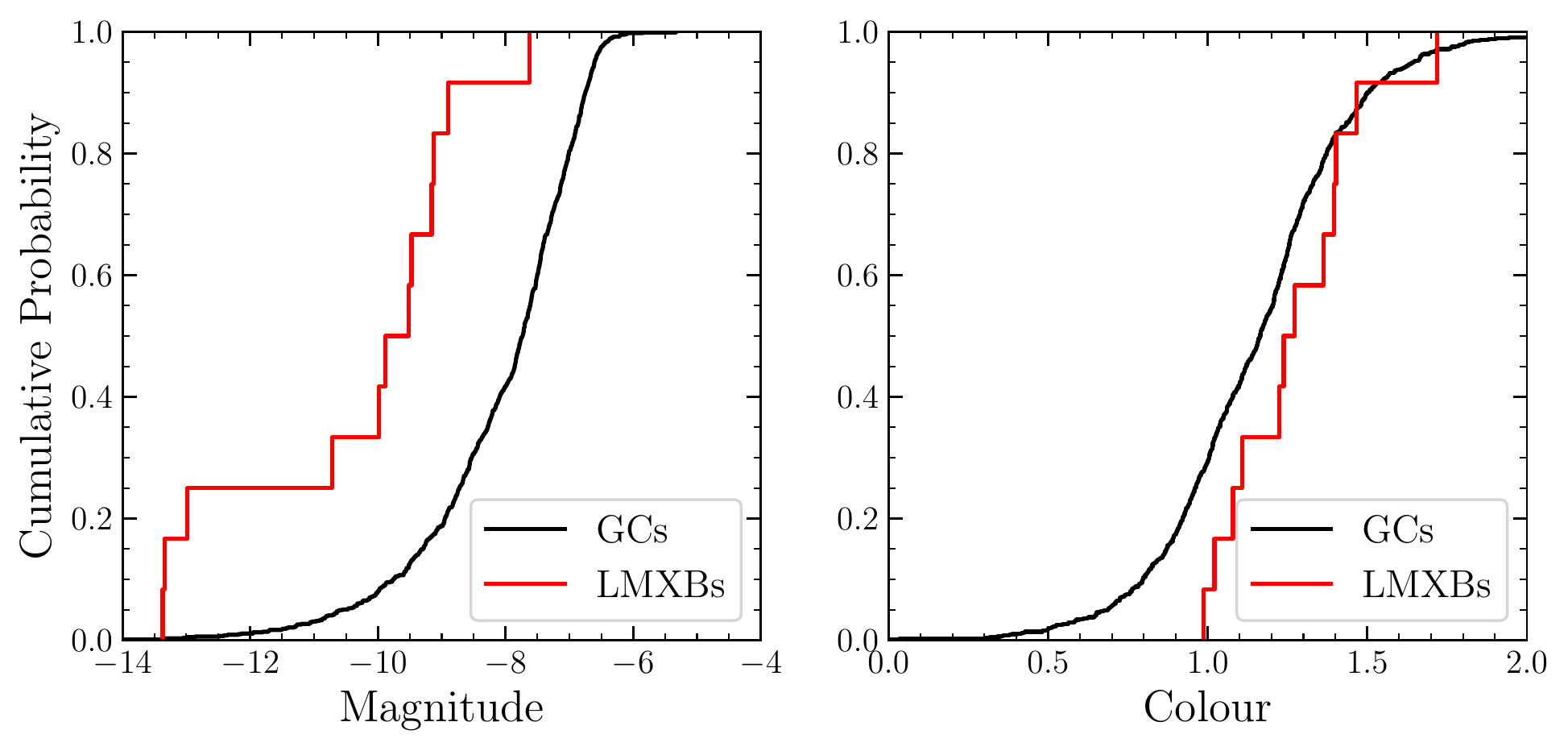}
\end{tabular}
\caption{Left:  Cumulative distribution functions for $z$ and $g-z$ for the ULX hosting clusters and non-ULX hosting clusters in the five galaxies (NGC 1316, NGC 1399, NGC 4472, NGC 4649 and M87 \citep{jordan09,2012ApJ...760...87S, peacock14,jordan16}. Right: Cumulative distribution functions for $z$ and $g-z$ for the NGC 1316 clusters and low X-ray luminosity hosting NGC 1316 clusters.} 
\label{fig:allulxs2}
\end{figure*}

Interestingly, while the NGC 1316 clusters that hosted lower luminosity X-ray sources showed a similar preference to be hosted by brighter clusters (see Figure \ref{fig:allulxs2}). The statistic value comparing the magnitude of the lower luminosity X-ray cluster hosts to the overall NGC 1316 cluster population is 12.6 and the significance level is $4.3\times10^{-7}$. However, the statistic value for optical colour is 1.52, with a significance level of 0.17, although this is could be due to the smaller sample size. % it is worth noting that of the twelve X-ray sources identified with clusters in this galaxy, eight of them have cluster colours redder than $g-z$=1.2.     

 %ib3n03050& 2011-07-30 &F160W ($H$)&2800

\subsection{Spatial Distribution of ULX Hosting Clusters}
\begin{figure}
\includegraphics[scale=0.4]{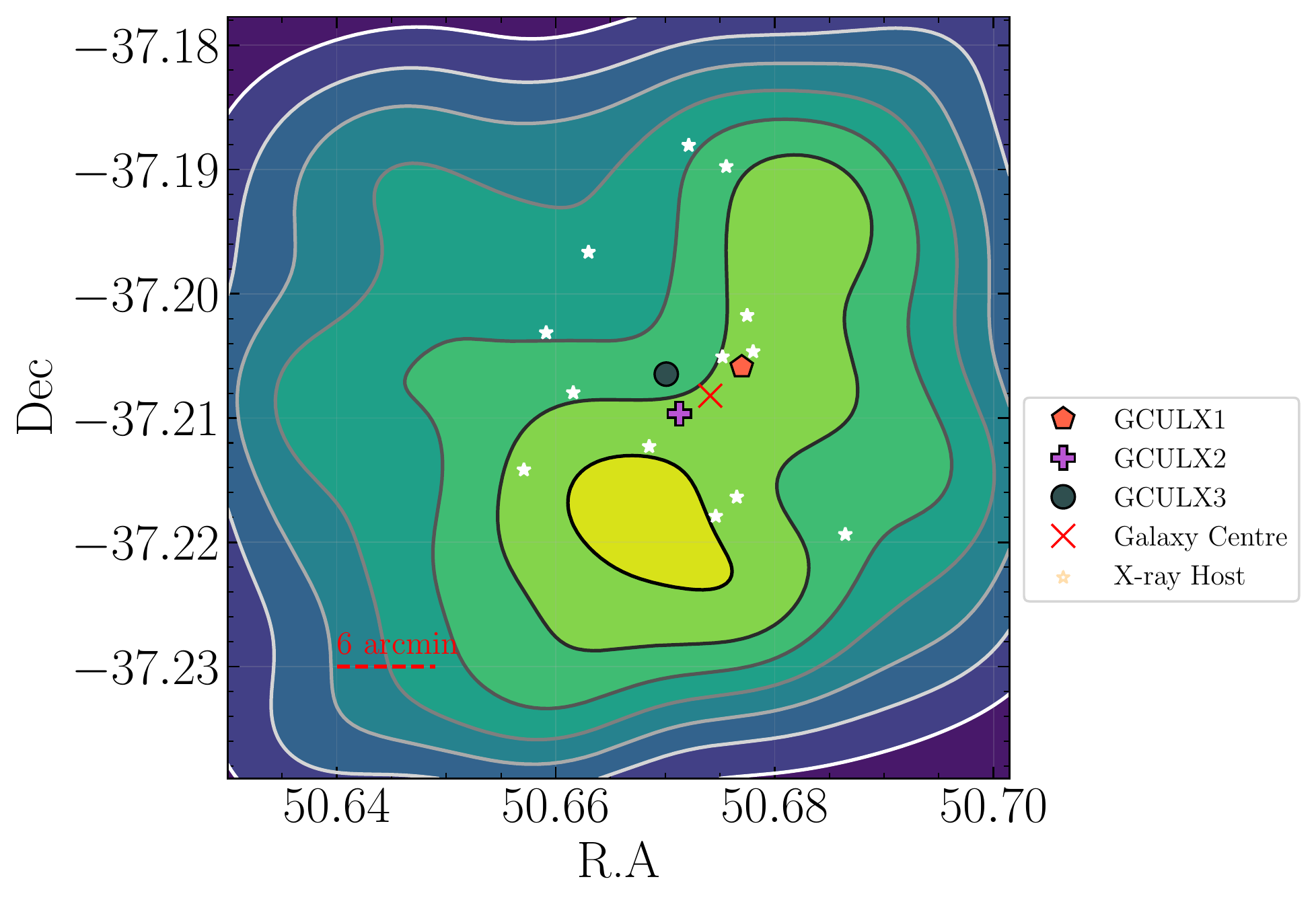}

\caption{Spatial distribution of globular clusters in NGC 1316, using globular clusters identified by \citet{jordan16}. We caution that the asymmetrical distribution of globular clusters may be an observational bias caused by obscuration from the dust lanes of the galaxy, although observational evidence for anisotropic globular cluster distribution has been observed in other galaxies such as NGC 4261 \citep{2005ApJ...634..272G}.  } 
\label{fig:raddist}
\end{figure}

\begin{figure}

\includegraphics[scale=0.45]{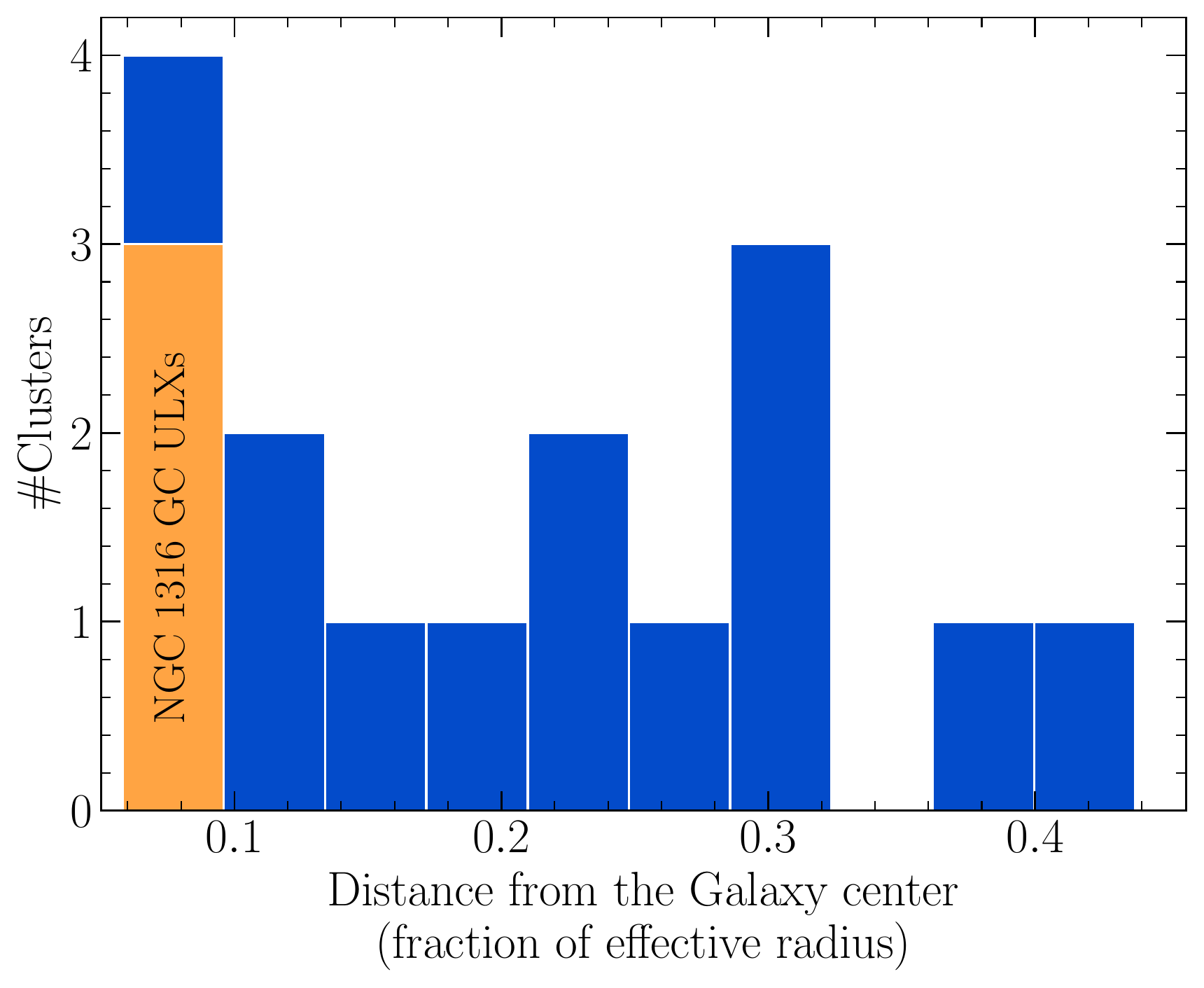}

\caption{Distance of NGC 1316's X-ray hosting clusters (Table \ref{sources}) from NGC 1316's galaxy centre as a function of effective radius (taken from the 2MASS catalog \citep{2006AJ....131.1163S}). } 
\label{fig:disthist2}
\end{figure}

 As shown in Figures \ref{fig:raddist}-\ref{fig:disthist2}, NGC 1316's globular cluster ULXs are all located near the galaxy centre, with the lower luminosity X-ray hosting clusters more spread out in the system.   

\begin{figure}

\includegraphics[scale=0.45]{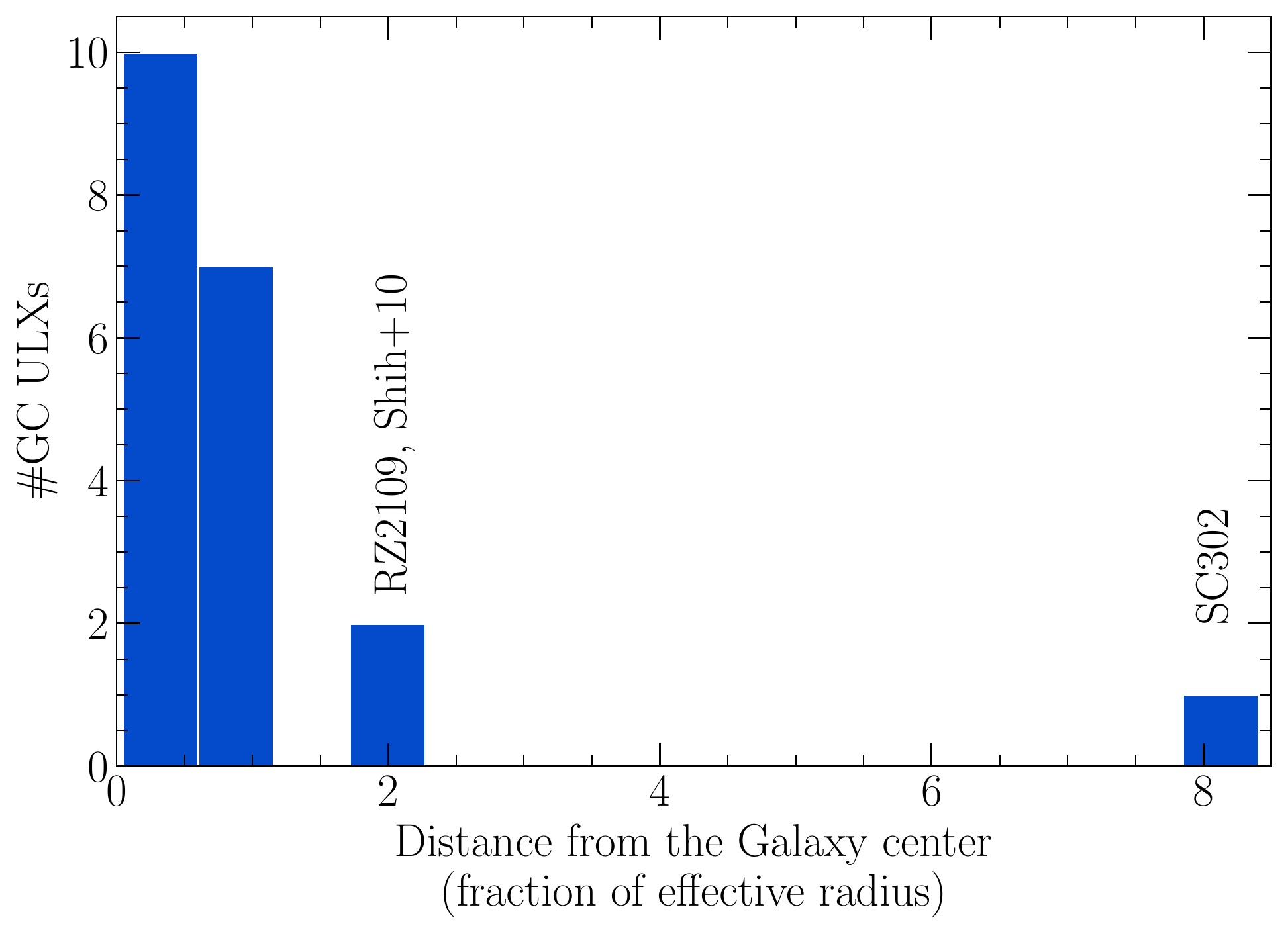}

\caption{Distance of ULX hosting clusters from their respective galaxy centres as a function of effective radius (taken from the 2MASS catalog \citealt{2006AJ....131.1163S}). The most distant ULX hosts are the NGC 1399 GC ULX observed by \citet{shih10}, the well-studied and highly variable source RZ2109 in NGC 4472 \citep{maccarone07}, and SC302 in M87 \citep{dage20}. } 
\label{fig:disthist}
\end{figure}

The full data-set of the now 20 discovered GC ULXs can help address the question of the spatial distribution of the ULX hosting clusters, and whether they are preferentially located near the galaxy centres or outskirts (see Figure \ref{fig:disthist}). 
%whether or not the ULX hosts follow the same spatial distribution of the overall cluster population, or
The three farthest ULXs were hosted by clusters over two effective radii from their galaxy centres. Little is known about the nature of the distant NGC 1399 source, as it has not been bright in X-rays since 2003 \citep{shih10}, but the NGC 4472 source, RZ2109, is one of the most well-studied GC ULXs, and is thought to be a stellar mass black hole accreting from a white dwarf. The host of SC302, the farthest GC ULX identified in M87, by contrast, has optical properties that are ambiguous as to whether the ULX is hosted by a globular cluster or perhaps a stripped nucleus \citep{dage20}. In the latter case, this may imply that the accretor of SC302 may be an intermediate mass black hole.  %Figures \ref{Fig:blah} and \ref{Fig:blah2} in the Appendix present the relative spatial locations and distances from the galaxy centres.

Given the current data, ULXs seem to be over-represented in the inner regions of the galaxy centre, although we caution that for some galaxies, there may not be complete lists of the globular clusters, or sufficient X-ray observations to cover the outer regions of all of these galaxies.

\subsection{ Power-Law Model Fits}
\begin{figure*}

\includegraphics[scale=0.5]{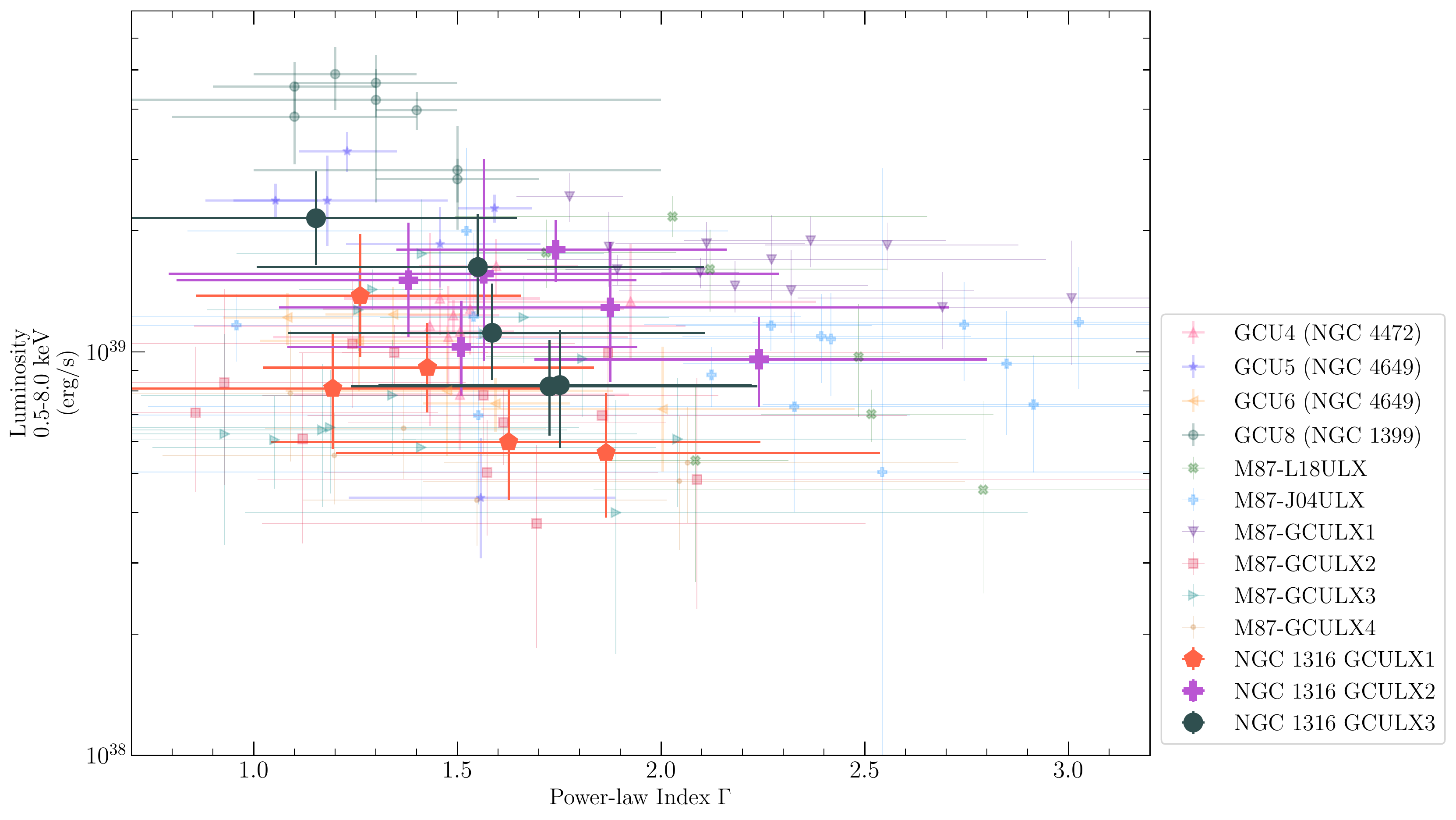}

\caption{$L_X$ versus $\Gamma$ for NGC 1316's GC ULX population compared to fit parameters for other known GC ULXs \citep{dage19a, dage20}. } 
\label{fig:plfits}
\end{figure*}

Using data from \citep{dage18, dage19a, dage20}, we can compare the power-law fits of the three NGC 1316 GC ULXs to the previously studied sample. Figure \ref{fig:plfits} displays the best fit power-law index ($\Gamma$) versus the X-ray luminosity (calculated assuming D=20.0 Mpc). The three NGC 1316 GC ULXs seem to follow the same trends as the previously studied sources. Interestingly, GCULX8 in NGC 1399 (where the X-ray luminosity was calculated with D=20.0 Mpc) is still an outlier compared to the rest of the sources, and is intrinsically much brighter, even compared to sources at a comparable distance.

\subsection{ Black-body Disk Model Fits}
\begin{figure*}

\includegraphics[scale=0.5]{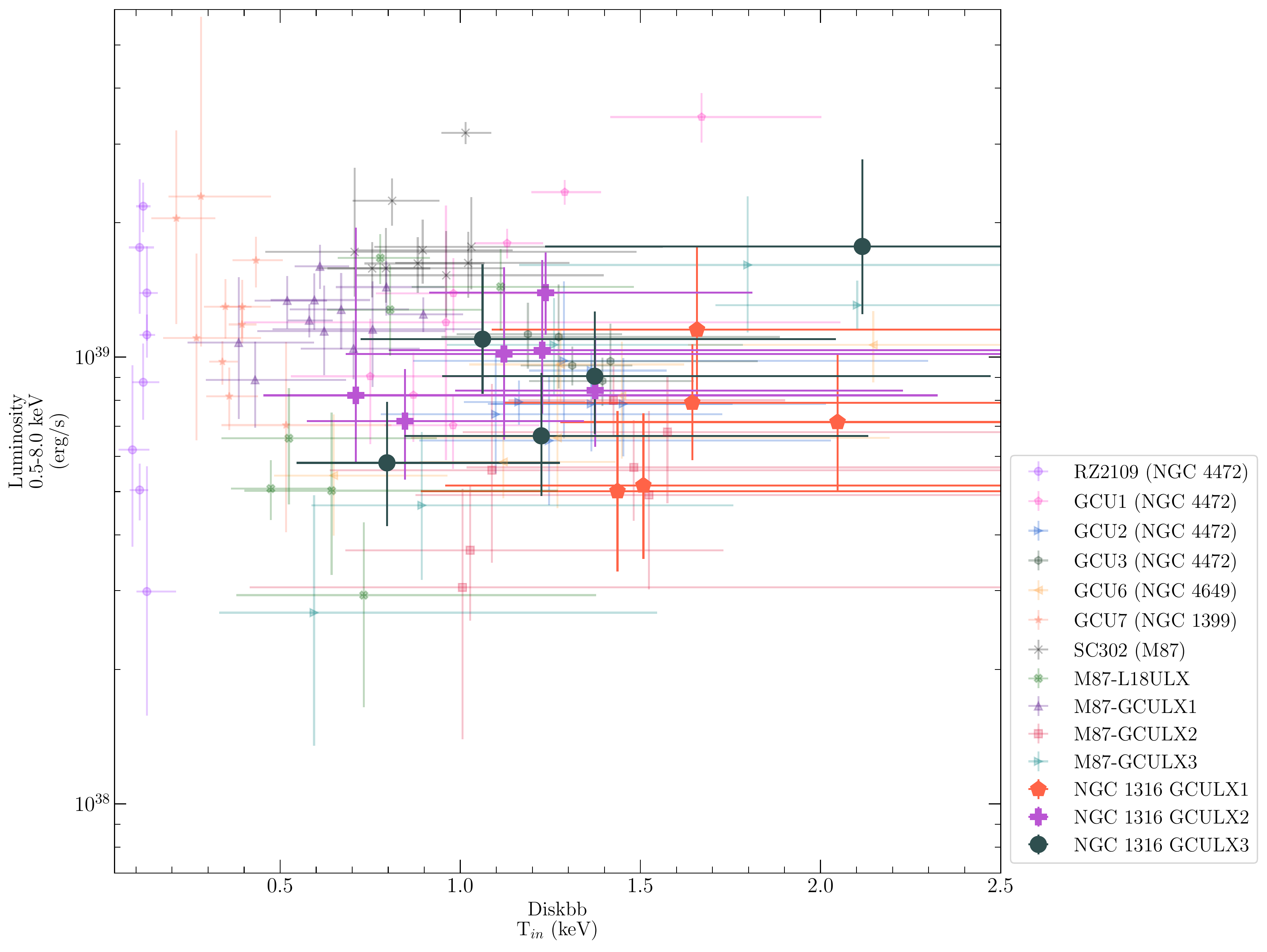}

\caption{$L_X$ versus $T_{in}$ for NGC 1316's GC ULX population compared to fit parameters for other known GC ULXs \citep{dage18, dage19a, dage20}. } 
\label{fig:disks}
\end{figure*}

Figure \ref{fig:disks} compares the changes in the measured inner disk temperature ($T_{in}$) to the X-ray luminosity. These sources also appear to follow the same trends as many of the other GC ULXs, where $L_X$ and kT have been established to be correlated in some cases \citep{dage19a}.%may be tightly correlated. 
Both RZ2109 in NGC 4472 and GCU7 in NGC 1399 behave differently than the rest of the GC ULXs, with both having low temperatures (kT $<$ 0.5 keV) which appear to be independent of the X-ray luminosity. These sources again highlight that the accretion astrophysics is not the same for all of the GC ULXs.

\section{Conclusions}
\label{sec:conc}

The three new GC ULXs identified in NGC 1316 bring the total known population of globular cluster ULXs to 20. Their behaviour in X-ray is neither as luminous as GCU8 in NGC 1399, nor as strange as RZ2109 in NGC 4472 or GCU7 in NGC 1399. Of the sources studied, the GC ULXs appear to still fall into the same three categories of spectral behaviour, (1.) the low kT sources like RZ2109 and GCU7, (2.) the ``intermediate" sources with high kT where both kT and $L_X$ vary, and (3.) the sources best fit by a power-law spectrum.  %sources that may follow a correlation between $L_X$ and kT

The intermediate sources in the second group, which have a wide range of best-fit spectral parameters, may be more comparable to well-studied bright Galactic X-ray binaries \citep{2016ApJS..222...15T}. Many of these sources behave like Galactic X-ray binaries, only with consistently brighter luminosities, or show behaviour similar to sources like GRS 1915  \citep{2016ApJS..222...15T} (although formed in an older population) which are bright, and only pass the Eddington limit for a 10 solar mass BH for a small fraction of the total observations.

The separate X-ray spectral behaviours of GC ULXs implies that the sources have different combinations of accretor masses/donor stars, and therefore also multiple paths to evolution and formation channels. Given that the X-ray binary formation process in a globular cluster is highly dynamic, and involves multiple interactions with other bodies in the cluster, as well as pair exchanges, it is no surprise that the binary makeup and accretion physics of these systems are also diverse.% Further evidence is given to this by the difference in ages of the NGC 1316 host clusters. 

Studies of the optical properties of the host clusters show that while the ULXs tend to be more often hosted by luminous clusters, evidence is finally emerging that the host clusters of ULXs are also preferentially redder. 

The nature of these sources prompt a number of open questions, both in terms of the high energy astrophysical phenomena that drive them, as well as the dynamical evolution and formation of these sources. However, given that they are potential tracers of BHs in GCs, they have important implications for the possible origin of merging black holes observed by LIGO, for which black holes in GCs are one of the leading theoretical possibilities \citep{2017ApJ...836L..26C,2021arXiv210107793R}.

It is currently unclear what properties of the early-type galaxies or their cluster systems will predict whether or not they host ULXs, aside from brighter clusters preferentially hosting ULXs. For instance, \citet{Brassington10} surveyed bright X-ray sources in NGC 3379, and while three X-ray sources that were found to coincide with globular clusters were luminous, they were still below the Eddington limit. %Thus, it is no surprise that the systems with a larger number of GCs (such as M87) host the most GC ULXs have a very large system of GCs.

NGC 1316 also is home to luminous, but not super-Eddington, X-ray point sources aligned with its globular clusters. Given the recent merger of the galaxy, and the two different ages of cluster populations, studies of the X-ray luminosity function of the low mass X-ray binary population are interesting possibilities for a future study.
\section*{Data Availability Statement}
The X-ray data in this article is publicly available through the \textit{Chandra} archive\footnote{\url{https://cda.harvard.edu/chaser/}} The optical data is available through MAST \footnote{\url{https://archive.stsci.edu/}}.
\section*{Acknowledgements}

This research has made use of data obtained from the Chandra Data Archive and the Chandra Source Catalog. We  also acknowledge use of NASA's Astrophysics Data System and arXiv. KCD and DH acknowledge funding from the Natural Sciences and Engineering Research Council of Canada (NSERC), the Canada Research Chairs (CRC) program, and the McGill Bob Wares Science Innovation Prospectors Fund. KCD acknowledges fellowship funding from the McGill Space Institute. AK and MBP acknowledge support from NASA through grant number GO-14738 from StSci. SEZ acknowledges support from Chandra grants GO9-20080X and AR0-21008B. 
KCD thanks Carol-Rose Little, Will Clarkson, Eric Koch, Dane Kleiner and Paolo Serra for their helpful discussion.
The following software and packages were used for analysis: \textsc{ciao}, software provided by the Chandra X-ray Center (CXC),   \textsc{heasoft} obtained from the High Energy Astrophysics Science Archive Research Center (HEASARC), a service of the Astrophysics Science Division at NASA/GSFC and of the Smithsonian Astrophysical Observatory's High Energy Astrophysics Division, SAOImage DS9, developed by Smithsonian Astrophysical Observatory,  \textsc{numpy} \citep{2011arXiv1102.1523V}, and  \textsc{matplotlib} \citep{2007CSE.....9...90H}. This publication makes use of data products from the Two Micron All Sky Survey, which is a joint project of the University of Massachusetts and the Infrared Processing and Analysis Center/California Institute of Technology, funded by the National Aeronautics and Space Administration and the National Science Foundation.

%%%%%%%%%%%%%%%%%%%%%%%%%%%%%%%%%%%%%%%%%%%%%%%%%%

%%%%%%%%%%%%%%%%%%%% REFERENCES %%%%%%%%%%%%%%%%%%

% The best way to enter references is to use BibTeX:

\bibliographystyle{mnras}
\bibliography{ngc1316gculx} % if your bibtex file is called example.bib

\bsp	% typesetting comment
\label{lastpage}
\end{document}